\documentclass[twocolumn,preprintnumbers,amsmath,amssymb,prl,superscriptaddress]{revtex4-2}
\usepackage{multirow}
\usepackage{amsfonts}
\usepackage[english]{babel}
\usepackage[T1]{fontenc}
\usepackage{times}
\usepackage{mathrsfs}
\usepackage{graphicx}
\usepackage{dcolumn}
\usepackage{bm}
\usepackage[colorlinks,bookmarks=true,citecolor=blue,linkcolor=red,urlcolor=blue]{hyperref}
\usepackage{epstopdf}

\begin{document}
\title{Biorthogonal-only Floquet Dynamical Quantum Phase Transitions}
\author{Jiangrong Wen}
\affiliation{Department of Physics and Chongqing Key Laboratory for Strongly Coupled Physics, Chongqing University, Chongqing 401331, China}
\author{Qidong Yuan}
\affiliation{Department of Physics and Chongqing Key Laboratory for Strongly Coupled Physics, Chongqing University, Chongqing 401331, China}
\author{Zi-Xiang Hu}
\email[Contact author: ]{zxhu@cqu.edu.cn}
\affiliation{Department of Physics and Chongqing Key Laboratory for Strongly Coupled Physics, Chongqing University, Chongqing 401331, China}
\author{Jian-Jun Dong}
\email[Contact author: ]{dongjianjun@cqu.edu.cn}
\affiliation{Department of Physics and Chongqing Key Laboratory for Strongly Coupled Physics, Chongqing University, Chongqing 401331, China}
\date{\today}

\begin{abstract}
Non-Hermitian dynamical quantum phase transitions (DQPTs) are intrinsically sensitive to the choice of inner product under nonunitary time evolution. Although the biorthogonal formulation based on associated states provides a normalized Loschmidt echo with a probabilistic interpretation, previous studies have found biorthogonal and self-normal DQPTs to occur in the same parameter regimes, suggesting that the two forms of dynamical criticality are concomitant. Here we demonstrate that this is not the case. In an exactly solvable periodically driven non-Hermitian Su-Schrieffer-Heeger chain, we uncover a finite biorthogonal-only Floquet DQPT regime, where the biorthogonal Loschmidt rate becomes nonanalytic while the self-normal Loschmidt rate remains smooth. The critical conditions are obtained analytically, showing that the onset of biorthogonal Floquet DQPTs is locked to the exceptional lines of the effective Floquet Hamiltonian, whereas self-normal criticality has no corresponding spectral boundary. Moreover, for each critical momentum, the biorthogonal DQPT exhibits a pair of critical times within every driving period, whereas the self-normal DQPT exhibits only one. Our results establish a fundamental distinction between biorthogonal and self-normal DQPTs, thereby opening a route toward new nonequilibrium quantum phenomena in non-Hermitian systems.
\end{abstract}

\maketitle
Non-Hermitian quantum systems host phenomena with no Hermitian counterparts, including exceptional points
\cite{Heiss2012JPA,Bergholtz2021RMP,Huang2021PRR}, the non-Hermitian skin effect
\cite{Lee2016PRL,Yao2018PRL,Zhang2020PRL,Guo2021PRL,XZhang2022AdvPhysX,Dash2026PRB},
and unconventional bulk-boundary correspondence
\cite{Kunst2018PRL,Edvardsson2019PRB,Yokomizo2019PRL,Helbig2020NatPhys,Cao2021PRB}.
These discoveries have greatly expanded the scope of quantum matter beyond the Hermitian paradigm.
While much progress has been made in understanding the spectral, topological, and steady-state properties of non-Hermitian systems, their far-from-equilibrium dynamics remains comparatively less explored.
Dynamical quantum phase transitions (DQPTs), signaled by nonanalyticities of the Loschmidt rate at critical times, provide a natural framework for characterizing nonequilibrium criticality
\cite{Heyl2013PRL,Andraschko2014PRB,Abeling2016PRB,Heyl2017PRB,Bhattacharya2017PRB,Kosior2018PRA,Sedlmayr2018PRB,Halimeh2018PRL,Halimeh2018PRB,Mera2018PRB,Heyl2018RPP,
Dong2019PRB,Bandyopadhyay2021PRL,Wong2022PRB,ShuChen2023PRB,GZYao2025PRA,Osborne2025PRRes,Mitra2026PRB,Shen2017PRL,Zhang2017Nature,PengXue2019PRL,HengFan2019PRAppl,Nie2020PRL,Tian2020PRL}.
Extending DQPTs to non-Hermitian systems is intrinsically subtle because nonunitary evolution allows inequivalent definitions of state overlap
\cite{Jing2024PRL,Zhou2018PRA,Zhou2021NJP,Mondal2022PRB,Mondal2023PRB,Naji2022PRA,XGao2025PRB,BMXu2026arXiv}.
Moreover, direct biorthogonal generalizations of the Loschmidt echo are not guaranteed to be real and nonnegative, and thus may lose their interpretation as return probabilities
\cite{HaoChen2025PRA,GaoyongSun2025PRB,FXLi2025PRBL,LibinFu2026CTP}.
This ambiguity reflects a fundamental feature of non-Hermitian quantum mechanics: the left and right eigenstates play distinct and generally inequivalent roles
\cite{CPSun1998PRE,Brody2014JPA,Ueda2020AdvPhys,Arkhipov2026PRRes}.

Building on the biorthogonal framework of non-Hermitian systems, we have recently proposed the notion of biorthogonal DQPTs formulated in terms of the time-evolved associated state \cite{Jing2024PRL}. This construction yields an automatically normalized biorthogonal Loschmidt echo and differs essentially from the conventional self-normal formulation, which is built solely from the right states. It has since been generalized to other models \cite{FuxiangLi2025PRB,Yao2026EPJB,Gu2026arXiv} and has also been experimentally explored in non-Hermitian quantum walks \cite{Zhamg2025LightSciAppl}. In the parameter regimes studied so far, however, biorthogonal and self-normal DQPTs have typically appeared together: both formulations predict dynamical singularities for the same system parameters, although the corresponding critical momenta or critical times may differ. Their distinction has therefore seemed to amount mainly to a shift of dynamical critical points, suggesting that the two types of DQPTs are necessarily concomitant. This observation raises a more fundamental question: can the choice of overlap structure determine not only where a dynamical singularity occurs but whether such a singularity appears at all?
In particular, can a biorthogonal DQPT occur in a parameter regime where no self-normal DQPT exists?

In this Letter, we answer this question affirmatively by studying an exactly solvable periodically driven non-Hermitian Su-Schrieffer-Heeger chain.
We identify a finite biorthogonal-only (B-only) Floquet DQPT regime, in which the biorthogonal Loschmidt rate becomes nonanalytic while the self-normal Loschmidt rate remains analytic for the same Hamiltonian parameters, initial state, and time evolution.
A complete parameter-space analysis reveals four distinct regimes: B-only, Both, self-normal-only (S-only), and None.
Our analytical solution shows that the two types of DQPTs are governed by different critical mechanisms.
Biorthogonal critical momenta are fixed by real-part gap closings of the time-independent effective Floquet Hamiltonian, so that exceptional lines mark the onset boundaries of biorthogonal DQPT regions.
By contrast, self-normal criticality is controlled by a right-state overlap condition and has no analogous spectral correspondence.
The two formulations also display distinct intraperiod temporal structures: each biorthogonal critical momentum gives rise to a pair of critical times within each driving period, whereas the self-normal construction yields only a single critical time per period.
These results establish biorthogonal-only Floquet DQPTs as a distinct form of nonequilibrium criticality and demonstrate that the overlap structure in non-Hermitian quantum mechanics can determine not merely the location, but the very existence of a dynamical phase transition.

\begin{figure}[ptb]
\begin{center}
\includegraphics[width=8.5cm]{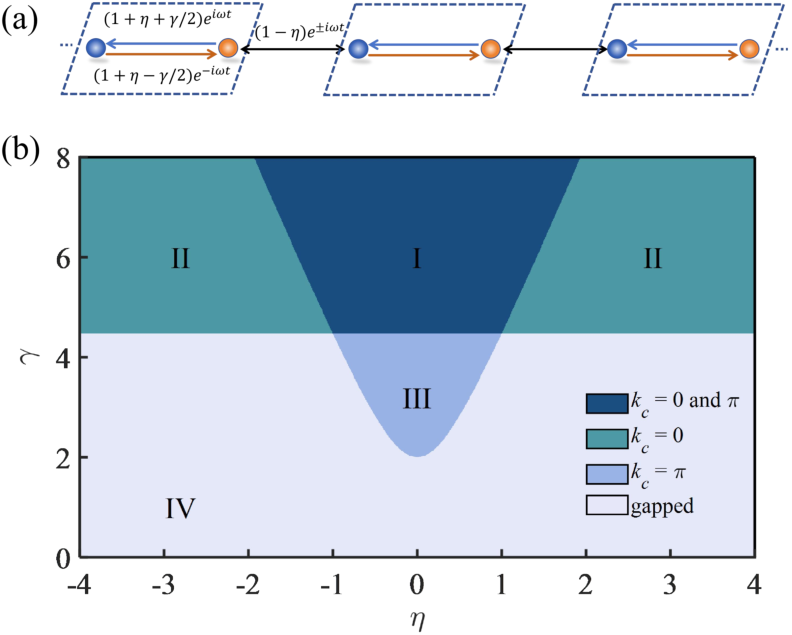}
\end{center}
\caption{ (a) Schematic of the periodically driven non-Hermitian SSH chain, with the dashed box indicates a unit cell.
(b) Four spectral regimes of $\mathcal{H}_{F}(k)$ in the $(\eta,\gamma)$ parameter plane, separated by exceptional lines corresponding to $k_c=0$ and $k_c=\pi$.
Region~I is real-part gapless at both $k_c=0$ and $k_c=\pi$; Regions~II and III are real-part gapless only at $k_c=0$ and $k_c=\pi$, respectively; Region~IV exhibits a finite real-part energy gap across the entire Brillouin zone.}
\label{fig1}
\end{figure}

To elucidate the origin of the biorthogonal-only Floquet DQPT regime, we consider an exactly solvable periodically driven non-Hermitian Su-Schrieffer-Heeger (SSH) chain, as illustrated in Fig.~\ref{fig1}(a),
\begin{align}
&H(t)=\sum_j \Bigg[ (1-\eta)e^{-i\omega t} c_{j+1,a}^{\dagger}c_{j,b} + \text{H.c.}\\
&+ \left(1+\eta-\frac{\gamma}{2}\right)e^{-i\omega t} c_{j,a}^{\dagger}c_{j,b}+ \left(1+\eta+\frac{\gamma}{2}\right)e^{i\omega t} c_{j,b}^{\dagger}c_{j,a} \Bigg]. \nonumber
\end{align}
Here, $c_{j,\alpha}^{\dagger}$ creates a particle on sublattice $\alpha=a,b$ in the $j$th unit cell.
The real parameters $\eta$ and $\gamma$ control the dimerization and intracell nonreciprocity, respectively, and $\omega$ is the driving frequency.
The Hamiltonian satisfies $H(t+T)=H(t)$ with $T=2\pi/\omega$.
We set $\hbar=1$ throughout and use $\omega=2$ and $N=8000$ unit cells in the numerical calculations. Such a driven non-Hermitian SSH chain can be realized in photonic resonator arrays or topolectrical circuits. The factors $e^{\pm i\omega t}$ correspond to time-dependent Peierls phases generated by electro-optic, acousto-optic, or parametric modulation, while the unequal intracell amplitudes can be engineered using direction-dependent gain/loss or active nonreciprocal circuit elements.
The two sublattices are encoded in two optical modes or circuit nodes, and the SSH dimerization is controlled by the relative strength of the intercell and intracell couplings.
These platforms also allow reconstruction of the Floquet evolution operator, providing access to both biorthogonal and self-normal Loschmidt rates.

Under periodic boundary conditions, we perform the Fourier transformation
$c_{j,\alpha}=N^{-1/2}\sum_k e^{ikj}c_{k,\alpha}$,
which gives $H(t)=\sum_k \Psi_k^{\dagger}\mathcal{H}(k,t)\Psi_k$ with
$\Psi_k^{\dagger}=(c_{k,a}^{\dagger},c_{k,b}^{\dagger})$.
The corresponding Bloch Hamiltonian reads
\begin{equation}
\mathcal{H}(k,t)=
\begin{pmatrix}
0 & (x_k-i y_k)e^{-i\omega t} \\
(x_k+i y_k)e^{i\omega t} & 0
\end{pmatrix},
\end{equation}
where
\begin{equation}
x_k=1+\eta+(1-\eta)\cos k,\quad y_k=(1-\eta)\sin k-i\frac{\gamma}{2}.
\end{equation}

The time-dependent Schr\"odinger equation $ i\partial_t|\Psi_k(t)\rangle = \mathcal{H}(k,t)|\Psi_k(t)\rangle$ can be solved exactly by introducing the rotating-frame transformation $ U_R(t) = e^{-i\omega t\sigma_z/2}$ and $ |\Phi_k(t)\rangle = U_R^{\dagger}(t)|\Psi_k(t)\rangle$ \cite{Yang2019PRB,Zamani2020PRB,Jafari2021PRA,Cai2022PRA,Naji2022PRB,Zamani2022PRB}.
The corresponding time-independent effective Hamiltonian is
\begin{equation}
\mathcal{H}_{F}(k)=U_{R}^{\dagger}(t)\mathcal{H}(k,t)U_{R}(t)-iU_{R}^{\dagger }(t)\partial_{t}U_{R}(t)=\mathbf{d}_{k}\cdot\bm{\sigma}.
\end{equation}
Here, $\mathbf{d}_k=(x_k,y_k,-\omega/2)$, while $\bm{\sigma}=(\sigma_x,\sigma_y,\sigma_z)$ denotes the vector of Pauli matrices.
The rotating-frame state therefore evolves as $ |\Phi_k(t)\rangle = e^{-i\mathcal{H}_{F}(k)t} |\Phi_k(0)\rangle$, and the laboratory-frame state is given by $ |\Psi_k(t)\rangle = U_R(t)e^{-i\mathcal{H}_{F}(k)t} |\Psi_k(0)\rangle$.

In the following, we characterize the spectral structure using the eigenvalues of $\mathcal{H}_{F}(k)$, $E_{k,\pm}=\pm E_k$, with $E_k=\sqrt{\mathbf{d}_k^2}$,
rather than the quasienergies $\varepsilon_{k,\pm}=\omega/2+E_{k,\pm}\pmod{\omega}$ folded into the first quasienergy Brillouin zone \cite{Holthaus2016JPB}.
This representation is more convenient for identifying real-part gap closings and exceptional lines.
At the two characteristic momenta, one has
\begin{equation}
E_{0}^{2} = \frac{16+\omega^{2}-\gamma^{2}}{4}, \quad E_{\pi}^{2} = \frac{16\eta^{2}+\omega^{2}-\gamma^{2}}{4}.
\end{equation}
The two eigenvalue branches coalesce at $E_k=0$ when $ \gamma^2=16+\omega^2$ for $k=0$ and $\gamma^2=16\eta^2+\omega^2$ for $k=\pi$.
These conditions define the exceptional lines associated with the $k=0$ and $k=\pi$ spectral closings and partition the parameter space into the four regimes shown in Fig.~\ref{fig1}(b).

To characterize these regimes, we define the real-part spectral gap as $ \Delta_{\mathrm{Re}}(k) = \left| \operatorname{Re}E_{k,+} - \operatorname{Re}E_{k,-} \right|$.
Since $ \operatorname{Im}(E_k^2) = -\gamma(1-\eta)\sin k$, a necessary condition for $E_k$ to become purely imaginary is $\sin k=0$ for generic $\gamma\neq0$ and $\eta\neq1$.
Thus, real-part gap closings can occur only at $k_c=0$ or $k_c=\pi$.
The additional condition $\operatorname{Re}(E_k^2)<0$ yields $\Delta_{\mathrm{Re}}(0)=0$ for $\gamma^2>16+\omega^2$, whereas $\Delta_{\mathrm{Re}}(\pi)=0$ for $\gamma^2>16\eta^2+\omega^2$.
Consequently, Region~I is real-part gapless at both $k_c=0$ and $k_c=\pi$, Region~II is real-part gapless only at $k_c=0$, Region~III only at $k_c=\pi$, while Region~IV remains real-part gapped throughout the Brillouin zone.

We next examine how the above spectral classification manifests itself in the biorthogonal dynamics.
Away from the exceptional lines, where $\mathcal{H}_{F}(k)$ is diagonalizable, its right and left eigenstates satisfy \cite{Brody2014JPA}
\begin{equation}
\mathcal{H}_{F}(k)|u_{k,\mu}\rangle = E_{k,\mu}|u_{k,\mu}\rangle, \quad \mathcal{H}_{F}^{\dagger}(k)|\widetilde{u}_{k,\mu}\rangle = E_{k,\mu}^{*}|\widetilde{u}_{k,\mu}\rangle,
\end{equation}
where $\mu=\pm$. They are chosen to satisfy the biorthogonal relations $ \langle\widetilde{u}_{k,\mu}|u_{k,\nu}\rangle = \delta_{\mu\nu}$ and $ \sum_{\mu=\pm} |u_{k,\mu}\rangle \langle\widetilde{u}_{k,\mu}| = \mathbb{I}$.
For a right state expanded in this basis as $ |\psi_k(t)\rangle = \sum_{\mu=\pm} c_{k,\mu}(t)|u_{k,\mu}\rangle$, the corresponding associated bra is $ \langle\widetilde{\psi}_k(t)| = \sum_{\mu=\pm} c_{k,\mu}^{*}(t) \langle\widetilde{u}_{k,\mu}|$.
For any nonzero state, the associated-state norm is therefore strictly positive, $ \langle\widetilde{\psi}_k(t)|\psi_k(t)\rangle = \sum_{\mu=\pm}|c_{k,\mu}(t)|^2>0$.
Within this associated-state framework, we introduce the normalized biorthogonal Loschmidt echo as \cite{Jing2024PRL}
\begin{equation}
L_{\mathrm{B}}(t) = \frac{\langle\widetilde{\Psi}(0)|\Psi(t)\rangle \langle\widetilde{\Psi}(t)|\Psi(0)\rangle }{\langle\widetilde{\Psi}(t)|\Psi(t)\rangle \langle\widetilde{\Psi}(0)|\Psi(0)\rangle },
\end{equation}
which satisfies $0\leq L_{\mathrm{B}}(t)\leq1$ and therefore admits a direct interpretation as a return probability.

We prepare the system in the product state occupying the minus branch of $\mathcal{H}_{F}(k)$ at each momentum, $ |\Psi(0)\rangle = \bigotimes_k |u_{k,-}\rangle$.
Since different momentum sectors evolve independently, the Loschmidt echo factorizes as $ L_{\mathrm{B}}(t) = \prod_k g_k^{\mathrm{B}}(t)$, with
\begin{equation}
g_k^{\mathrm{B}}(t) = \frac{ \langle\widetilde{u}_{k,-}|u_{k,-}(t)\rangle \langle\widetilde{u}_{k,-}(t)|u_{k,-}\rangle }{\langle\widetilde{u}_{k,-}(t)|u_{k,-}(t)\rangle },
\end{equation}
where $\langle\widetilde{u}_{k,-}|u_{k,-}\rangle=1$ has been used.
The time-evolved right state is
\begin{equation}
|u_{k,-}(t)\rangle=U_{R}(t)e^{-i\mathcal{H}_{F}(k)t}|u_{k,-}\rangle=\sum_{\mu=\pm}c_{k,\mu}(t)|u_{k,\mu}\rangle,
\end{equation}
with $ c_{k,\mu}(t) = e^{-iE_{k,-}t} \langle\widetilde{u}_{k,\mu}| U_R(t) |u_{k,-}\rangle$.
Consequently, the single-mode echo $g_{k}^{\mathrm{B}}$ can be written as
\begin{equation}
g_{k}^{\mathrm{B}}(t)=\frac{|c_{k,-}(t)|^2}{\displaystyle\sum_{\mu=\pm}|c_{k,\mu}(t)|^2} = \frac{\left| \langle\widetilde{u}_{k,-}| U_R(t)
|u_{k,-}\rangle \right|^2 }{ \displaystyle \sum_{\mu=\pm} \left| \langle\widetilde{u}_{k,\mu}| U_R(t) |u_{k,-}\rangle \right|^2}. \label{eq:simplified-mode-echo}
\end{equation}
The common factor $\left|e^{-iE_{k,-}t}\right|^2$ cancels upon normalization, making $0\leq g_k^{\mathrm{B}}(t)\leq1$ explicit.
The biorthogonal Loschmidt rate is defined as
\begin{equation}
\lambda_{\mathrm{B}}(t) = -\lim_{N\rightarrow\infty} \frac{1}{N}\ln L_{\mathrm{B}}(t) = -\int_{-\pi}^{\pi} \frac{dk}{2\pi} \ln g_k^{\mathrm{B}}(t).
\end{equation}
Since the denominator of Eq.~(\ref{eq:simplified-mode-echo}) is strictly positive for a nonzero state, $g_k^{\mathrm{B}}(t)$ vanishes if and only if $ \langle\widetilde{u}_{k_c,-}|U_R(t_c)|u_{k_c,-}\rangle=0$.
The existence of such a real solution $(k_c,t_c)$ signals a biorthogonal Floquet DQPT and generically gives rise to a nonanalytic cusp in the rate function $\lambda_{\mathrm{B}}(t)$.

Solving the condition $ \langle\widetilde{u}_{k_c,-}| U_R(t_c) |u_{k_c,-}\rangle = 0 $ yields the critical-momentum condition $ \operatorname{Re}E_{k_c,-} = 0$ and $E_{k_c,-}\neq0$, where the second condition excludes the exceptional lines, on which the two eigenstates coalesce.
For each critical momentum in the real-part-gapless regime, two critical times occur within the $n$th driving period:
\begin{equation}
t_{c,1}^{(n)} = nT+\tau_{k_c}, \quad t_{c,2}^{(n)}= (n+1)T-\tau_{k_c},
\end{equation}
where $ \tau_{k_{c}}=\frac{2}{\omega}\arctan\left( \frac{1}{\omega} \sqrt{\gamma^{2}-\omega ^{2}-4x_{k_{c}}^{2}} \right) $.
The two critical times are equidistant from the midpoint of each driving period, $ \frac{ t_{c,1}^{(n)}+t_{c,2}^{(n)} }{2} = \left(n+\frac{1}{2}\right)T$.
Since $E_{k,\pm}=\pm E_k$, the condition $\operatorname{Re}E_{k_c,-}=0$ is equivalent to $\Delta_{\mathrm{Re}}(k_c)=0$.
Thus, a real-part gap closing in the spectrum of $\mathcal{H}_{F}(k)$ determines the existence of a critical momentum of biorthogonal DQPT, while the exceptional lines mark the corresponding onset boundaries.

\begin{figure}[ptb]
\centering
\includegraphics[width=8.5cm]{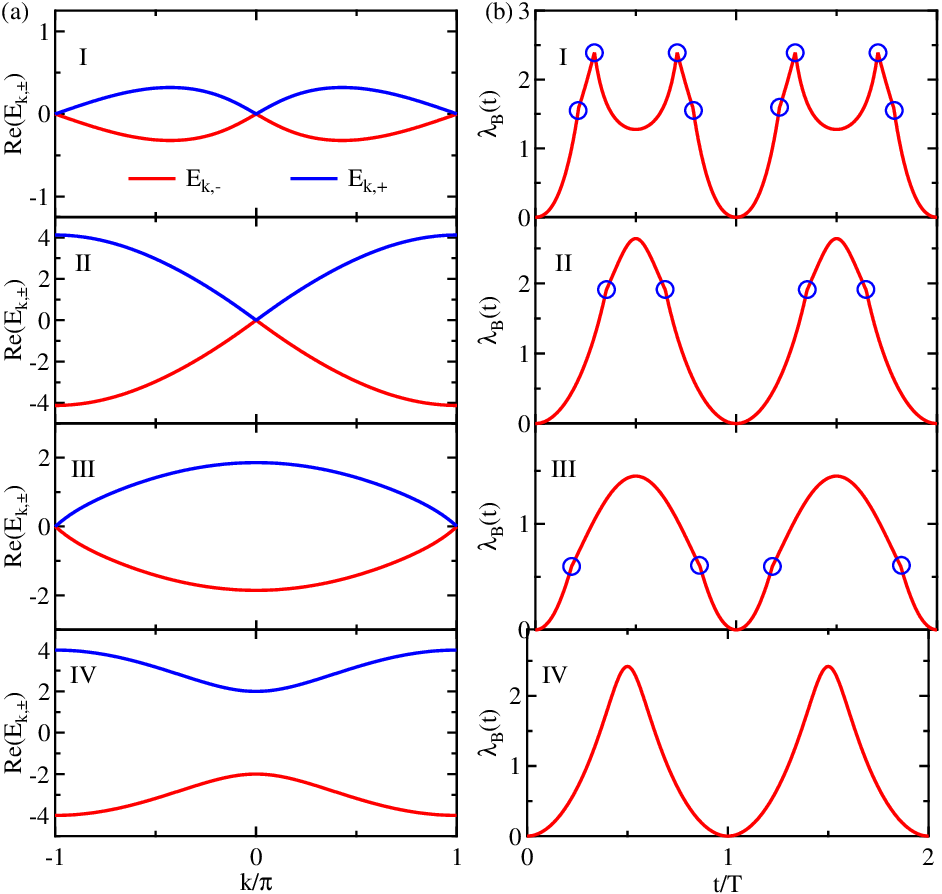}
\caption{ (a) Real parts of the dispersions $\operatorname{Re}E_{k,\pm}$ in Regions~I--IV.
(b) Corresponding biorthogonal Loschmidt rates $\lambda_{\mathrm{B}}(t)$. Circles mark the critical points.
The parameters are $(\eta,\gamma)=(0.85,4.75)$, $(2.5,6)$, $(0.2,2.5)$, and $(2.0,2.0)$ for Regions~I--IV, respectively.}
\label{fig2}
\end{figure}

As shown in Fig.~\ref{fig2}, Region~I supports two critical momenta, $k_c=0$ and $k_c=\pi$. Each critical momentum generates a pair of critical times within every driving period, yielding two pairs of critical times in Region~I.
As a comparison, Regions~II and III support only one critical momentum, $k_c=0$ and $k_c=\pi$, respectively, and therefore exhibit only one pair of critical times per period.
In Region~IV, the real-part spectrum remains gapped throughout the Brillouin zone; hence no critical momentum exists and no biorthogonal Floquet DQPT occurs.

The above results establish a direct connection between real-part gap closings in the spectrum of the effective time-independent Hamiltonian and biorthogonal Floquet DQPTs.
We next ask whether biorthogonal criticality is necessarily accompanied by a self-normal Floquet DQPT for the same system parameters.
To address this question, we consider the self-normal Loschmidt echo $L_{\mathrm{S}}(t) = |\langle\Psi(0)|\Psi(t)\rangle|^2$ with a renormalized factor \cite{Zhou2018PRA}.
For the product initial state introduced above, the independent momentum sectors give
\begin{equation}
L_{\mathrm{S}}(t) = \prod_k \frac{ |\langle u_{k,-}|u_{k,-}(t)\rangle|^2 }{ \langle u_{k,-}|u_{k,-}\rangle \langle u_{k,-}(t)|u_{k,-}(t)\rangle } \equiv \prod_k g_k^{\mathrm{S}}(t).
\label{eq:self-normal-mode-echo}
\end{equation}
The corresponding self-normal Loschmidt rate is $\lambda_{\mathrm{S}}(t) =-\int_{-\pi}^{\pi}\frac{dk}{2\pi} \ln g_k^{\mathrm{S}}(t)$.
A self-normal Floquet DQPT occurs if and only if $\langle u_{k_c,-}|u_{k_c,-}(t_c)\rangle=0$.
Solving this condition gives the critical times $t_{c,\mathrm{S}}^{(n)}=(n+\frac{1}{2})T$.
Thus, self-normal criticality exhibits a single critical time at the midpoint of each driving period, independently of the critical momentum. By contrast, each biorthogonal critical momentum generates a pair of critical times within every driving period.

\begin{figure}[ptb]
\centering
\includegraphics[width=8.5cm]{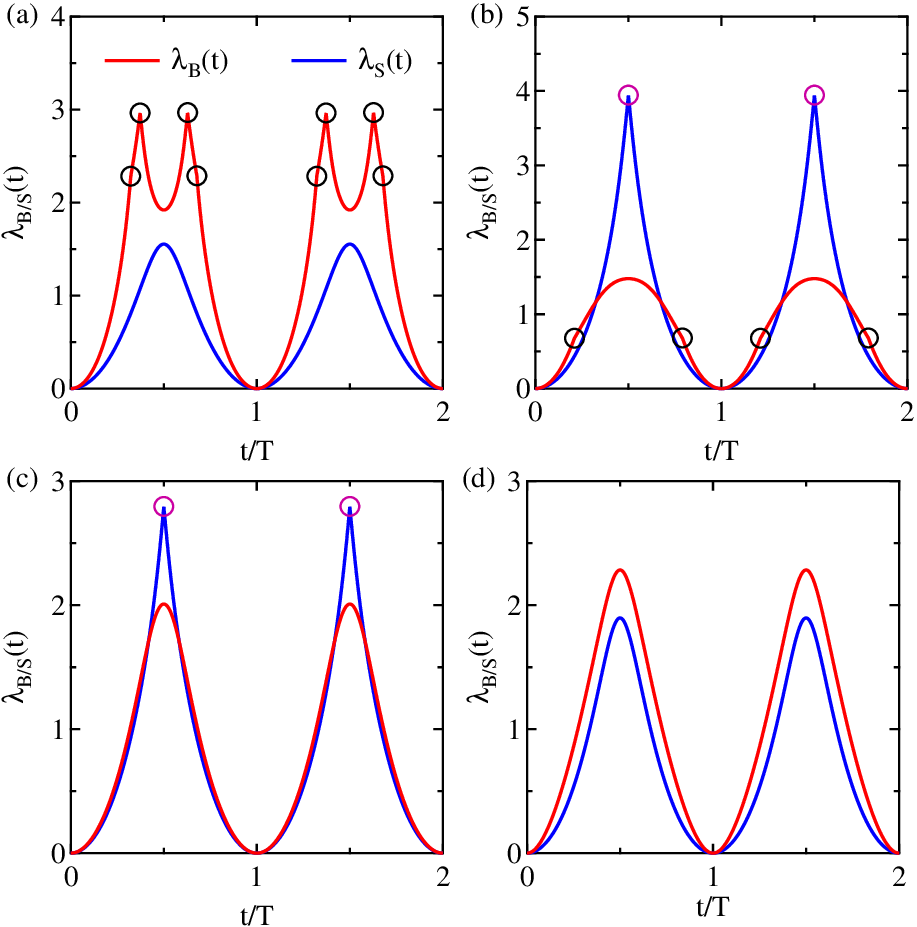}
\caption{ Representative comparisons of the biorthogonal and self-normal Loschmidt rates for the same Hamiltonian parameters, initial state, and time evolution. Circles mark the corresponding critical points.
(a) Biorthogonal-only: $\lambda_{\mathrm{B}}(t)$ develops nonanalytic cusps, whereas $\lambda_{\mathrm{S}}(t)$ remains smooth.
(b) Both: both rates exhibit nonanalyticities.
(c) Self-normal-only: nonanalyticities occur only in $\lambda_{\mathrm{S}}(t)$.
(d) Neither: both rates remain analytic.
The parameters $(\eta,\gamma)$ are $(0.5,5.5)$, $(-0.15,2.6)$, $(-1.5,2.5)$, and $(2.0,4.4)$ for (a)--(d), respectively.
}
\label{fig3}
\end{figure}

Figure~\ref{fig3} illustrates all four possible dynamical behaviors.
Most importantly, Fig.~\ref{fig3}(a) provides a direct example of a biorthogonal-only Floquet DQPT, in which the biorthogonal rate develops nonanalytic cusps while the self-normal rate remains analytic under the same initial state and evolution.
The remaining panels illustrate the three other scenarios: coexistence of both types of Floquet DQPTs, self-normal-only criticality, and a regime where neither form of DQPT occurs.
Therefore, the difference between the two formulations is more than just a shift of the dynamical critical points: the selected state-overlap structure can decide whether a Floquet DQPT occurs at all.

\begin{figure}[ptb]
\centering
\includegraphics[width=8.5cm]{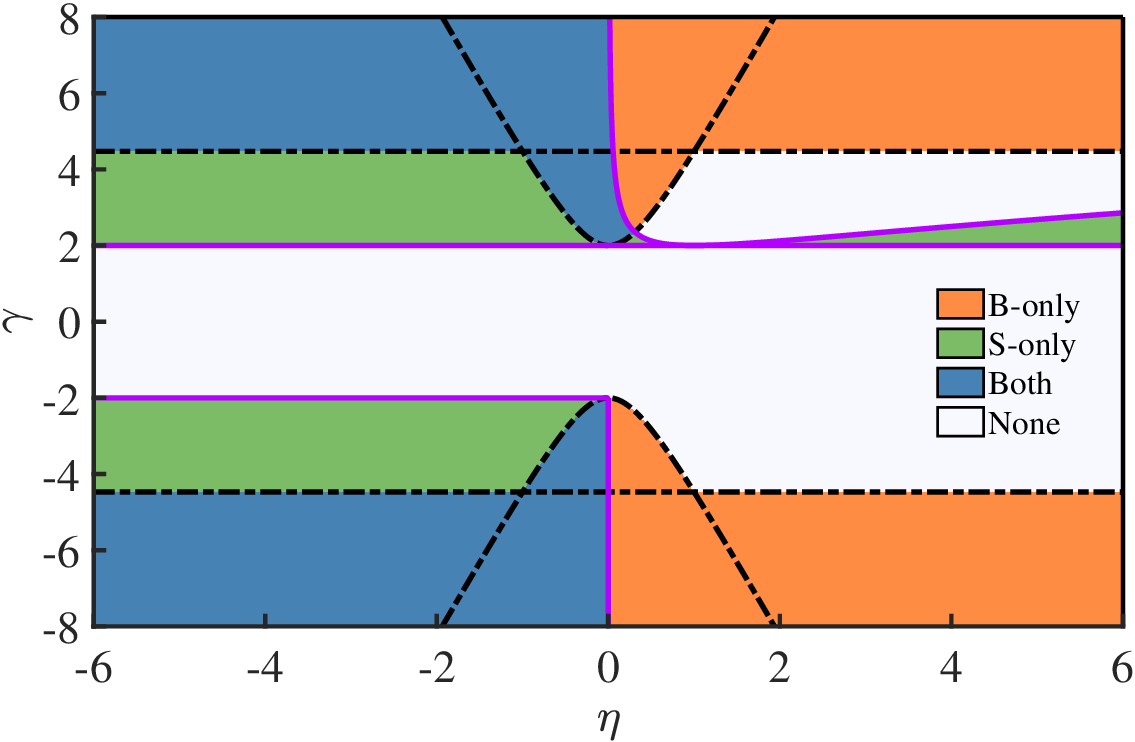}
\caption{ Global Floquet-DQPT diagram for the biorthogonal and self-normal formulations in the $(\eta,\gamma)$ plane.
The four regimes, B-only, Both, S-only, and None, indicate whether a Floquet DQPT occurs only in the biorthogonal formulation, in both formulations, only in the self-normal formulation, or in neither formulation, respectively.
The black dashed curves mark the exceptional lines, which coincide with the onset boundaries of biorthogonal Floquet DQPTs.
The purple solid curves mark the self-normal DQPT boundaries determined by Eq.~(\ref{eq:self-normal-condition}).}
\label{fig4}
\end{figure}

To determine whether these distinct behaviors persist over finite parameter ranges, we derive the conditions under which $\langle u_{k_c,-}|u_{k_c,-}(t_c)\rangle=0$ admits a real critical solution associated with the initially chosen $E_{k,-}$ branch. This yields
\begin{equation}
1\leq\frac{\gamma}{\omega}\leq\frac{1+\eta}{2\sqrt{\eta}}\quad(\eta >0),\quad\left(  \frac{\gamma}{\omega}\right)  ^{2}\geq1\quad(\eta\leq0).
\label{eq:self-normal-condition}
\end{equation}
These conditions define the parameter region supporting self-normal Floquet DQPTs.
Unlike the biorthogonal boundaries, which are determined by exceptional lines and are symmetric under $\gamma\rightarrow-\gamma$, the self-normal critical region depends additionally on the initially chosen eigenvalue branch.
Reversing $\gamma$ therefore does not generally preserve the admissible critical solution, giving rise to the asymmetry seen in Fig.~\ref{fig4}.
This branch dependence also shows that self-normal criticality is not tied to the same spectral boundaries.
By jointly applying the biorthogonal and self-normal critical criteria, Fig.~\ref{fig4} divides the parameter space into four distinct regimes: B-only, Both, S-only, and None.
Most importantly, the B-only regime spans a finite area instead of consisting of isolated points, demonstrating that biorthogonal-only Floquet criticality is not a fine-tuned phenomenon but rather persists across a broad range of system parameters.

In summary, we have uncovered a finite biorthogonal-only Floquet DQPT regime in an exactly solvable, periodically driven non-Hermitian SSH chain, in which a biorthogonal DQPT occurs whereas no self-normal DQPT occurs under the same preparation and evolution.
We have shown that biorthogonal criticality is tied to real-part gap closings of the effective Hamiltonian, with its onset boundaries coinciding with exceptional lines, whereas self-normal criticality exhibits no analogous spectral correspondence.
Furthermore, within each driving period, biorthogonal critical times always appear in pairs, whereas self-normal criticality exhibits only a single critical time.
This establishes that biorthogonal dynamical criticality is not only quantitatively distinct from its self-normal counterpart, but can also exist in a regime where the latter is entirely absent. Our results therefore show that the overlap structure is a decisive ingredient in non-Hermitian Floquet dynamical criticality.

More broadly, the biorthogonal structure of non-Hermitian eigenspaces provides an additional degree of freedom for constructing dynamical overlaps and characterizing nonequilibrium evolution \cite{Sun2022FP,Tang2022EPL,Lu2026PRB}.
Unlike in Hermitian systems, where the dual state is fixed by Hermitian conjugation, left and right eigenstates in non-Hermitian systems generally carry independent information.
Utilizing this additional structure to uncover novel nonequilibrium phenomena is therefore one of the central challenges in non-Hermitian physics.
The finite B-only regime uncovered in this work serves as a concrete illustration that the biorthogonal framework can give rise to dynamical criticality that does not appear in descriptions relying exclusively on right states. Generalizing this approach to a broader class of non-Hermitian models, driving schemes, and many-body configurations could reveal additional nonequilibrium behavior that has no counterpart in right-state-only treatments.

This work is supported by the National Natural Science Foundation of China Grants No. 12204075 and No. 12547101, the Natural Science Foundation of Chongqing Grant No. CSTB2023NSCQ-MSX0953 and No. CSTB2024YCJH-KYXM0064.
Z.~H was supported by the National Natural Science Foundation of China Grant No.12474140 and the Fundamental Research Funds for the Central Universities Grant No. 2025CDJ-IAISYB-029.

\end{document}